\documentclass{emulateapj}

\newcommand{\nid}{\noindent}
\newcommand{\ul}{\underline}

\newcommand{\ds}{\displaystyle}

\newcommand{\beq}{\begin{equation}} 
\newcommand{\eeq}{\end{equation}} 

\usepackage{graphicx} 
\usepackage{natbib}

\begin{document}

\shorttitle{Pulsar timing and space interferometry}
\shortauthors{Spallicci}

\title{ On the complementarity of pulsar timing and space laser interferometry for the individual detection of supermassive black hole binaries}
\author {Alessandro D.A.M. Spallicci}
\affil{{Universit\'e d'Orl\'eans, Observatoire des Sciences de l'Univers en r\'egion Centre, UMS 3116} \\ 
{LPC2E-CNRS UMR { 7328}, 3A Av. Recherche Scientifique, 45071 Orl\'eans, France}}
\email{spallicci@cnrs-orleans.fr}

\date{9 January 2013}


\begin{abstract}
Gravitational waves coming from Super Massive Black Hole Binaries (SMBHBs) are targeted by both Pulsar Timing Array (PTA) and Space Laser Interferometry (SLI). The possibility of a { single} SMBHB being tracked first by PTA, through inspiral, and later by SLI, up to  
{ merger}  and { ring down}, has been previously suggested.    
Although the bounding parameters are drawn by the current PTA or the upcoming Square Kilometer Array (SKA), and by the New Gravitational Observatory (NGO), derived from the Laser Interferometer Space Antenna (LISA), { this paper also addresses} sequential detection beyond specific project constraints. We consider PTA-SKA, { which is} sensitive from $10^{-9}$ to 
$p\cdot10^{-7}$ Hz ($p=4,~~8$), and SLI, { which operates} from 
$s\cdot10^{-5}$ up to $1$ Hz ($s = 1,~~3 $). A SMBHB in the range $2\cdot 10^8 - 2 \cdot 10^9~M_\odot$ (the masses are normalised to a 
$(1+z)$ factor, { the red shift lying between  $z = 0.2$ and $z=1.5$}) moves from the PTA-SKA to the SLI { band over a period} ranging from two months to fifty years. 
By combining 
three Super Massive Black Hole (SMBH)-host relations with three accretion prescriptions, nine astrophysical scenarios are formed. They are then related to three levels of pulsar timing residuals ($50$, $5$, $1$ ns), generating twenty-seven cases. 
For residuals of $1$ ns, sequential detection probability will never be better than $4.7 \cdot 10^{-4}$ y$^{-2}$ or $3.3 \cdot 10^{-6}$ y$^{-2}$(per year to { merger}  and per year of survey), according to { the} best and worst astrophysical scenarios, respectively; { put differently this means} one sequential detection every $46$ or $550$ years for an equivalent { maximum} time to { merger} and duration of the survey.  
The chances of sequential detection are further reduced by increasing values of the $s$ parameter (they vanish for $s = 10$) and of the SLI noise, and by decreasing values of the remnant spin.
The spread in the predictions diminishes when { timing precision is improved or the SLI low frequency cut-off is lowered}.  So while transit times and the SLI Signal to Noise Ratio (SNR) may be adequate, the likelihood of sequential detection is { severely hampered} by the current estimates on the number - just an handful - of { individual inspirals observable} by PTA-SKA, and to a lesser extent by the { wide gap} between the pulsar timing and space interferometry bands, and by the severe requirements on pulsar timing residuals. Optimisation of future operational scenarios for SKA and SLI is briefly dealt with, since a detection of even a single event would be of paramount importance for the understanding of SMBHBs and of the astrophysical processes connected to their formation and evolution.   

\end{abstract}



\section{Introduction}

At different paces, Earth-based and space gravitational wave detectors and observatories are already - or will { soon} be - operating. They { aim} to cover different parts of the gravitational spectrum, targeting an extremely large variety of astrophysical sources. Nonetheless, large gaps will show between the sensitive frequency bands. One such gap { is that} between the Pulsar Timing Array (PTA) - in the future the Square Kilometer Array (SKA) - and the Space Laser Interferometry (SLI) bands. SLI may materialise as the New Gravitational Observatory (NGO), { derived from} the Laser Interferometer Space Antenna (LISA). Though their bands lie apart, PTA-SKA and SLI may monitor the same type of source, namely a Super Massive Black Hole Binary (SMBHB) at different evolutionary stages.  
In this paper, we explore under what circumstances { a single} SMBHB may be viewed first by PTA-SKA and later by SLI.

{ The motivation for the study lies} in the opportunities offered by the analysis of the same sources with different instruments ({ radio astronomy} and interferometry), and at different relativistic regimes (inspiral, coalescence, { merger}  and ring-down). If pulsar timing { were to} provide mass and spin parameters, the latter could be counter-checked by laser interferometry. Likewise, a binary system could be examined, {\it vis--\`a--vis} the presence of matter, gas and other bodies, at different phases of { its} evolution.      
 

\subsection{{ Observation} by pulsar timing}

 In the frequency band between $10^{-9}$ Hz and some fraction of $10^{-6}$ Hz, PTA offers the unique chance to observe gravitational radiation. Beyond the { observation} of a stochastic background, the { challenge of} { observing a single} SMBHB by PTA has recently received growing attention. Simulations { concur} { in predicting } a scenario wherein some SMBHBs stick out of the stochastic { signal} floor of unresolved SMBHBs. 
{ The observation of an individual SMBHB} provides opportunities for { new measurements} in general relativity { and new perspectives for the scientific community}. 

{ The state} of the art on timing residual precision\footnote{The timing residuals are computed from the phase difference between the observed time of arrival (ToA) and the predicted ToA, based on the current model parameters.} from current  PTAs\footnote{PPTA {\scriptsize http://www.atnf.csiro.au/research/pulsar/ppta}, {\scriptsize EPTA http://www.epta.eu.org/}, NANOGrav{\scriptsize http://nanograv.org/}.} { lies} in the $100-50$ ns domain{ , while} improvement by SKA\footnote{SKA {\scriptsize http://www.skatelescope.org/}.} down to $10-1$ ns is expected (Liu et al. 2011). 

Investigations\footnote{In the following cited studies, two simplifying { hypotheses} have been adopted: i) binaries are on circular orbits; ii) the mergers are gravitational wave driven.} on individual sources aim to recover { physical parameters such as the spins and the masses} of, and the distance to, an SMBHB (Sesana \& Vecchio 2010) using the gravitational { wave front} curvature (Deng \& Finn 2011), or the Pulsar term\footnote{The 'Earth term' is the gravitational wave strain at the Earth at the time when the pulse is received. The 'Pulsar term' is the strain at the pulsar at the time when the pulse is emitted. An SMBHB produces two quasi-monochromatic components in PTA residuals, and likely of different frequencies as the SMBHB evolves.} (Corbin \& Cornish 2010; Lee et al. 2011; Mingarelli et al. 2012). From these studies it emerges that the distance to a pulsar is important for individual { SMBHBs}, { unlike} statistical background observation (Jenet et al. 2004; Sesana et al. 2009; Finn \& Lommen 2010). { Finn and Lommen (2010) and Pitkin (2012) analysed} bursts coming from different sources including individual { SMBHBs}. 

Burt et al. (2011) suggest { refraining from searching for} new pulsars, unless close to an { existing cluster} of good pulsars; instead, they recommend { the allocation} of more { observation} time to already low-noise pulsars. 

Turning to specific { observation} targets, searches have so far produced negative results: Jenet et al. (2004) { found no evidence of the emission} of gravitational waves by a { supposed} SMBHB in 3C 66B; { neither did} Lommen \& Backer (2001), who were seeking evidence for an SMBHB in Sgr A*. 

Finally, Yardley et al. (2010) describe the { observations} used to produce the sensitivity curves for the Parkes { { radio telescope}}, and propose a method for detecting significant sinusoids in { PTA}.  

\subsection{Detection by space laser interferometry}

Gravitational wave detection in space was first proposed by means of a small sized interferometer on-board a single satellite (Grassi-Strini et al. 1979), before shaping into a triangular satellite configuration (Bertotti 1984; Faller et al. 1985,1989). 
{ The} LISA Pathfinder (Antonucci et al. 2012) is deemed an important step { towards technological} maturity. 

{ NGO\footnote{\scriptsize http://sci.esa.int/ngo} (ESA 2011), which { is derived from the previous LISA proposal}\footnote{\scriptsize http://sci.esa.int/lisa}, is a space project} designed to measure gravitational radiation over a broad band at low frequencies, where the Universe is richly populated by strong sources of gravitational
waves, including SMBHBs. 
NGO plans to trace the formation, growth and merger history of SMBHs during different epochs, 
measuring spin and masses, with an unprecedented precision, 
often where the Universe is blind with our current
electromagnetic techniques (ESA 2011). 
In fundamental physics, different tests on general relativity, including the no-hair theorem and the dynamics in strong-field, and on alternative theories, will be feasible with { SLI}.
NGO { (Amaro--Seoane et al. 2012a,b)} implies a shift to higher frequencies of the sensitivity band, as compared to LISA. 


\subsection{Sequential detection}

Pitkin et al. (2008) { first proposed} sequential detection, but { it appeared} necessary to improve and update   their initial work for several reasons. { We begin by ascertaining} that the total SMBHB mass is generally larger than the fifty million solar masses considered by Pitkin et al. (2008).  New { analyses} are carried out herein. First, we examine various astrophysical scenarios { combining} SMBH-host relations and accretion processes, { and} dry and wet mergers. Second, we span a large range of residuals ($50$--$1$ ns). Third, we estimate the number of events and the probability of sequential detection, building our investigation upon the recent { statistical} findings of individual detection by PTA-SKA, { which were} not available at the time of the work { by} Pitkin et al. (2008). Fourth, we present the Signal to Noise Ratio (SNR) of SLI for the sources concerned. 


\subsection{Structure of the paper} 

Section 2 is devoted to the computation of transit times (how long it takes for a binary to switch { bands}) from the PTA-SKA to the SLI band for a range of SMBHB masses. 
In the same section, the ringing frequencies are computed for { the} (stationary and rotating) { SMBHB remnants}.

Section 3 is the core of the { paper, in which we} study the impact of astrophysical, observational and detection constraints separately, one step after the other, steadily building up our analysis. Details of the models are described by Sesana et al. (2009), to which { the reader is referred for further information}. Super Massive Black Hole (SMBH)-host relation models { are then combined} with accretion prescription models, and different values of timing residuals { are considered}. We determine first the { maximum} number of sequential detections within a given time to { merger}, { assuming} that SLI would catch all the sources { that were previously observed by PTA-SKA}. { We then} consider the impact of the SLI low cut-off frequency, and of the spin of the remnants.  
Then for NGO, we compute the SNR at different values of the redshift $z$, and we finally adapt our previous estimates for sequential detection.   

Section 4 sums up the conclusions; the appendix attempts to sketch some of the operational scenarios that may lie ahead.  

We refer to the total mass of the binary $M$, normalised to $1 + z$, where $z$ is the { red shift} (Hughes 2002). The mass enters in the orbit evolution equation with the { time scale} $Gm/c^3$, the { time scale} being red shifted. The consequence\footnote{Petiteau et al. (2011),  discuss the feasibility of breaking the degeneracy with electromagnetic counterparts and propose enforcing statistical consistency.} is that a binary at $z=1.5$ and of mass $2\cdot 10^8~M_\odot$ is equivalent to a binary at $z = 0.2$ and of mass $4.17\cdot 10^8~M_\odot$. Herein, given the range of $z$, the factor $(1+z)$ introduces an uncertainty of less than $2.25$ on the mass determination. 

We use the terms of observation and detection when referring to PTA-SKA and SLI, respectively, { while} sequential detection implies both observation by PTA-SKA and detection by SLI. Finally, the term `event' refers mostly to astrophysical phenomena, or it is used whenever a specific labelling { is not intended}.


\section{Transit time and ringing}

Although notable advances in the two-body problem have been achieved (Blanchet et al. 2011), the foundations laid by  
Peters \& Mathews (1963) { and} Peters (1964) (PM) still suffice for the description of the relativistic binaries { for the present} analysis.
The main assumptions (Pierro \& Pinto 1996; Pierro et al. 2001) of the PM model  are: (i) point { masses}, (ii) weak field, (iii) slow motion, and (iv) adiabatic evolution (negligible change { in} the orbital parameters over each orbit).  

In the PM model, the time it takes a circularised binary, of equal masses $m_1 = m_2= m$ and total mass $m_1+m_2=M$, to evolve between two frequencies is given by\footnote{Equation (\ref{transittime}) is obtained by { integrating} equation (4) in Forward \& Berman (1967), as Peters and Mathews { did not} write an expression for the frequency evolution.} 
 
\beq
t_{\mathit t} = t_{\mathit p} - t_{\mathit s}=  \kappa M^{-5/3} \left( f_{\mathit p}^{-8/3} - f_{\mathit s}^{-8/3}\right ), 
\label{transittime}
\eeq

\nid where $t_{\mathit t}$ is the time of transit from the PTA-SKA band (exited at $t_{\mathit p}$ time) to the SLI band (entered at 
$t_{\mathit s}$ time); { $f_{\mathit p}$ and $f_{\mathit s}$ are the { PTA-SKA} high frequency and the SLI low frequency cut-offs, respectively;} { the numerical coefficient is given by

\[
\kappa = \ds 5 \cdot 2^{-35/9}\pi^{-8/3} \left(\ds\frac{G}{c^3} \right )^{-5/3}~, 
\]
$G$ being the constant of gravitation, $c$ the speed of light.

The $f_{\mathit p}$ cut-off { is determined by the interval between observations with the radio telescope}.  A daily allocation would bring the cut-off to $10^{-5}$ Hz. We have taken a semi-conservative stand by setting two values for $f_{\mathit p}$, namely $4 \cdot 10^{-7}$ Hz and $8 \cdot 10^{-7}$ Hz. If $f_{\mathit s} \gg f_{\mathit p}$, the value of $f_{\mathit s}$ becomes irrelevant in the computation of the transit time. In this regard, a shift towards higher frequencies of the band of SLI is not consequential. Conversely, the chances for sequential detection are strongly dependent { on even} a slight shift of $f_{\mathit s}$, and further, if the space interferometer has a modest sensitivity at low frequencies, the sequential detection may be easily missed, { see} Section 3. Figure \ref{fig1n.eps} shows the transit time.

The coalescence frequency, the frequency $f_c$ at which the post-Newtonian expansion of the inspiral ceases to be accurate at around $6M$ 
(Hughes 2002), and the ring-down frequency $f_r$ (Echeverria 1989; Hughes 2002) are given by

\beq
f_c \simeq { 4} \cdot 10^{-6}\frac{10^9~ M_\odot}{M}~Hz,
\eeq

\beq
f_r \simeq 3.2 \cdot 10^{-5}\frac{10^9~ M_\odot}{\eta M} [1 - 0.63 (1 - a)^{3/10} ]~Hz.
\label{eche}
\eeq

The ring-down frequencies { fall} within the band of some of the SLI configurations, especially for a high spin Kerr parameter $a$, see Figure \ref{fig2nn.eps}. The parameter $\eta = 0.94$ takes into account the emission of gravitational radiation\footnote{ According to numerical simulations, the mass radiated in gravitational waves is $M_{rad}/M = 1 - M_{fin}/M = 5–-7 \cdot 10^{-2}$, with $M = M_1 + M_2$  being the total mass (Rezzolla 2009).}.

Transit time and ring-down frequencies appear compatible { with} sequential detection. The shortest transit time is determined primarily by the $f_{\mathit p}$ frequency, while the longest transit time { is determined by the interval chosen between observation by PTA-SKA, and detection} by SLI (see Appendix).

\begin{figure} 
\includegraphics[width=8.5cm]{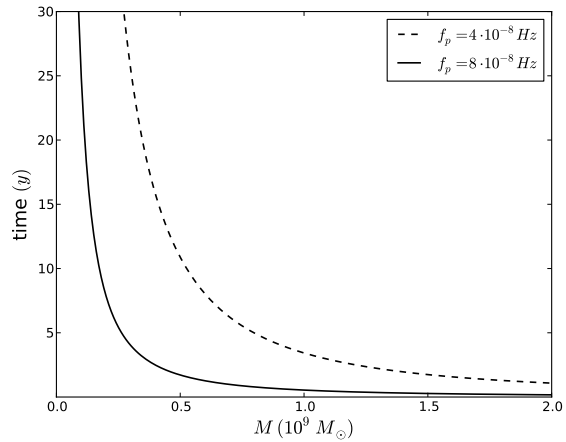}
\caption{Time of transit (years) as { a} function of the SMBHB total normalised mass, from { the} PTA-SKA sensitivity band ($f_{\mathit p} = 4 \cdot 10^{-7}$ Hz, dashed line or $8 \cdot 10^{-7}$ Hz, { solid} line) to { the} SLI sensitivity band ($f_{\mathit s} = 2 \cdot 10^{-5}$ Hz).}
\label{fig1n.eps}
\end{figure}

\begin{figure} 
\includegraphics[width=8.5cm]{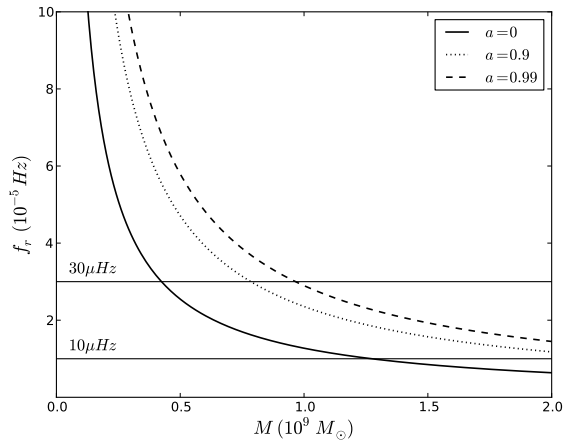}
\caption{Frequency (Hz) of ringing as { a} function of the { final SMBH} total normalised mass, for three  dimensionless spin Kerr parameter $a$ values: { $0$} ({ solid} line), $0.9$ { (dotted line)}, and $0.99$ (dashed line).  The horizontal lines correspond to two different values of $f_{\mathit s}$, i.e. $10$ or $30$ $\mu$Hz. }
\label{fig2nn.eps}
\end{figure}


\section{Observation and detection rates and probabilities, signal to noise ratio}

In this section, we combine three Super Massive Black Hole (SMBH)-host relations with three accretion prescriptions. The nine scenarios 
are further coupled with three different values of timing { residuals,} producing twenty-seven cases. For each of { these} cases, 
we retain the SMBHBs { that generate} a timing residual above a given threshold 
($50$, $5$, $1$ ns). There are two { time scales}: { $t_{\mathit r}$, which} fixes the { maximum} time to { merger} { that}  we intend to observe, { and $\Delta t_{\mathit l}$ which} refers to the SLI survey duration, {\it i.e.} the SLI mission lifetime (see Appendix). 

Firstly, see Table \ref{tab1}, we compute the number of observations of those SMBHBs which are {\it individually} detectable by PTA-SKA, and which are up to $10$ years away from { merger}: these values include those SMBHBs which { would not} enter any {\it currently considered} SLI band, as they are too massive. Conversely, if we imagine that the PTA-SKA and SLI bands are contiguous, the values in Table \ref{tab1} { take on} the meaning of {\it ideal} upper values.   
Then, see Tables \ref{tab2},\ref{tab3}, we derive a sub-set of the previous ensemble of SMBHBs whose remnants {\it all} enter the SLI band, for  two different values of the SLI low cut-off frequencies, and for different remnant spins. { An {\it optimal} sequential detection rate corresponds to this sub-set}, having { assumed} so far a noiseless interferometer. 
Finally for the same sources, see Table \ref{tab4}, we focus on NGO and compute the SNR 
at different values of the { red shift} $z$  in the range $0.2-1.5$, where the observable SMBHBs are expected to be (Sesana et al. 2009). { Lastly, we comment on} our previous estimates for sequential detection, in light of the characteristics (noise and cut-off) of the NGO project.  

The number of events are expressed in different units. We use { $n_{10,10}$} units, that is the $n$ number of events in $10$ years; more precisely the number of SMBHBs merging within $10$ years for a { survey lasting $10$ years}. The conversion to the number of events per year to { merger}  and per year of survey, { $N_{y^{-2}}$}, implies a division of { $n_{10,10}$} by a factor { of} 100.
{\it Ad hoc} factors have to be applied, when considering that the SLI lifetime is shorter than the PTA-SKA duration survey (see Appendix). An example of the third type of unit may be represented by { $n_{3,20}$, meaning $3$ years for $\Delta t_{\mathit l}$, and $20$ years for $\Delta t_{\mathit PTA-SKA}$, the duration of the PTA-SKA survey}.  
Two simplifying identities appear justified, namely equating: i) $t_{\mathit r}$ to $\Delta t_{\mathit PTA-SKA}$; ii) the end of the survey by PTA-SKA to the end of the SLI mission (see Appendix). 


\subsection{PTA-SKA observation rates (Ideal sequential detection)}

{ Concerning individual} { observation} solely by PTA, Sesana \& Vecchio (2010) estimated { that only a handful of observable binaries exist}, including those at { the} $5$ ns effective noise level; the number of resolvable systems
quickly drops if the timing precision degrades to, {\it e.g.}, $50$ ns. This result confirms that of Sesana et al. (2009), who analysed a wide range of population models and indicated 5-to-15 individual sources having residuals larger than the stochastic background\footnote{For a complementary approach see Boyle \& Pen 2010.}. The residuals are between $2$ and $60$ ns in the frequency range $2\cdot 10^{-8}
- 10^{-7}$ Hz.  Further, they identified most of the individually resolvable SMBHB sources as having a mass larger than $0.5\cdot 10^9~ M_\odot$ and lying at a { red shift}$ 0.2 < z < 1.5 $.

These findings are based on a { statistical sample of merging massive galaxies generated} from the
online Millennium database\footnote{Millenium {\scriptsize http://www.g-vo.org/Millennium}.} built by Springel et al. (2005). 
The Millennium simulation covers a comoving volume of $(500/h_{100})^3$~Mpc$^3$~$(h_{100} =
H_0/100$~km s$^{-1}$ Mpc$^{-1}$ is the normalised Hubble parameter). 
Sesana and co-workers populated the merging galaxies with central SMBHs according to different models. For each model, the expected distribution of bright individual sources
and the associated timing residuals were computed. 

The following procedure has been adopted. Out of the Millenium database, for every cell of the ($m_1, m_2,z$) distribution, 
we determine the radiating frequency of those SMBHBs which are i) in an orbital phase at most $t_{\mathit r}$ time before { merger}; ii) radiating at a frequency below $f_{\mathit p}$; iii) generating a timing residual larger than a given threshold. We compute the number of SMBHBs { per} for unit time and { per} frequency bin (of size equal to $1/t_{\mathit r}$) such that: ${\cal N}(f) = \int 1/t_{\mathit r}~d{\cal N}(f)/df$.  
Afterwards, we integrate $d^4{\cal N}/dm_1~dm_2~dz~dt$ along $t_{\mathit r}$ and obtain the number of sources for a given bin. Finally, the results are summed up for all bins, masses and { red shifts}. The outcome is the number of sources, individually observable by PTA-SKA, merging in $t_{\mathit r}$ time. 

{ This} observation probability depends on the details of the merging SMBHB population,
and specifically on the number of { coalescing} binaries and 
on their mass. Dependence on other intrinsic binary parameters such 
as spin and eccentricity may be negligible, though recent studies show that $1.5$ post-Newtonian terms, including spin-orbits effects may be detected (Mingarelli et al. 2012). Indeed, such 
systems, in the PTA-SKA band, will be far enough from { merger for} 
spin-orbit and spin-spin terms { not to affect our analysis significantly}; but will be close enough to { merger}, such that the residual 
eccentricity is likely to be smaller than $ 0.1$.

Since binaries producing timing residuals in this frequency range are
{ assumed} to be extremely massive systems, the rates will depend on
the high mass end of the Super Massive Black Hole (SMBH) mass function. 
This is an important point
since this is precisely where the two most popular
Massive Black Hole (MBH) mass predictors give inconsistent results.
MBH masses inferred by the M-$\sigma$ relation (Tremaine et al. 2002; Gultekin et al. 2009; Graham et al. 2011) can be indeed up to an
order of magnitude lower with respect to their M-bulge inferred
counterparts (H\"aring \& Rix 2004; Tundo et al. 2007; Lauer et al. 2007; Gultekin et al. 2009). 

We have tested nine different models, implementing three different
SMBH-host relations and three different accretion prescriptions. On this { basis}, we have 
built our catalogue of merging SMBHs starting from the
Millennium Run. Nine cases result { from} the cross-combinations.  

The { SMBH-host relations explored}, including intrinsic scattering (Sesana et al. 2009), are configured in three 
types.
\begin{itemize}
\item {A. M-$\sigma$ relation according to Tremaine et al. (2002). Following their  
relation, SMBH $\propto \sigma^4$, implies that SMBHs with 
masses larger than $10^9 M_\odot$ are extremely rare.}
\item{B. M-$\sigma$ relation according to Gultekin et al. (2009). This relation
shows a steeper dependence with $\sigma$, SMBH $\propto \sigma^{4.24}$; the intrinsic scattering is larger than in Tremaine et 
al. 2002, predicting more massive binaries.}
\item{C. M-bulge relation, { again} according to Gultekin et al. (2009). The M-bulge
relation generally predicts higher SMBH masses for a given
galaxy host, see, {\it e.g.}, H\"aring \& Rix (2004), Lauer et al. (2007), Tundo et al. (2007).
In this case, we have SMBH up to $10^{10} M_\odot$.}
\end{itemize}

Accretion prescriptions (Sesana et al. 2009) are grouped in three classes.
\begin{itemize}
\item{a. Accretion occurs onto the SMBH's remnant, meaning that the two
merging SMBHs are undermassive with respect to the selected
SMBH-host relation.}
\item{b. Accretion occurs onto the primary SMBH before coalescence, 
meaning that the total mass of the two merging systems follows
the SMBH-host relation, but the mass ratio is usually quite 
high (implying { a} weaker gravitational wave signal).}
\item{c. Accretion occurs onto both SMBHs before coalescence, 
meaning that the total mass of the two merging systems { follows}
the SMBH-host relation, and the mass ratio of the merging SMBH
is closer to unity (the most favorable situation for gravitational wave detection).}
\end{itemize}

\begin{table}
\centering{
\caption{ PTA-SKA observation rates \\(Ideal sequential detection upper values)}
\label{tab1}
\begin{tabular}{@{}lccc}
\hline
Model & $ 50$ ns & $5$ ns & $1$ ns\\
\hline
Aa & $3.5 ~\cdot10^{-9}$ & $3.4 ~\cdot10^{-6}$ & $3.3 ~\cdot10^{-4}$  \\
Ab & $1.4 ~\cdot10^{-8}$ & $1.8 ~\cdot10^{-5}$ & $1.5 ~\cdot10^{-3}$  \\
Ac & $9.5 ~\cdot10^{-8}$ & $1.2 ~\cdot10^{-4}$ & $5.5 ~\cdot10^{-3}$  \\
Ba & $8.7 ~\cdot10^{-9}$ & $2.1 ~\cdot10^{-5}$ & $9.1 ~\cdot10^{-4}$  \\
Bb & $2.3 ~\cdot10^{-7}$ & $1.4 ~\cdot10^{-4}$ & $4.0 ~\cdot10^{-3}$  \\
Bc & $3.9 ~\cdot10^{-6}$ & $1.3 ~\cdot10^{-3}$ & $1.6 ~\cdot10^{-2}$  \\
Ca & $8.2 ~\cdot10^{-7}$ & $3.3 ~\cdot10^{-4}$ & $4.3 ~\cdot10^{-3}$  \\
Cb & $1.5 ~\cdot10^{-5}$ & $1.6 ~\cdot10^{-3}$ & $1.5 ~\cdot10^{-2}$  \\
Cc & $3.5 ~\cdot10^{-4}$ & $9.7 ~\cdot10^{-3}$ & $4.7 ~\cdot10^{-2}$  \\
\hline
\end{tabular}}
\vskip5pt
\parbox[s]{8.7cm}{
{\footnotesize NOTE.-- { Total number of PTA-SKA} observations in { $10$} years ({ $n_{10,10}$} units)
for SMBHBs individually observable and generating a timing residual above a given threshold 
($50$, $5$, $1$ ns), merging in less than 10 years, and not necessarily destined to enter the SLI band, if too massive. 
Three models of SMBH-host relations and three accretion types have been explored, producing nine cross-combinations, each evaluated for three timing residual options. The resolvable SMBHBs are mostly situated at a { red shift} $0.2 < z < 1.5 $.}
}

\end{table}

The combination of the above assumptions { encompasses} a large 
range of SMBH mass functions, bracketing the observations.
The most pessimistic model - Aa - { predicts that hardly any SMBHBs with a total mass larger than $10^9M_\odot$ will coalesce}, whilst the most optimistic { model} - Cc - delivers 
coalescences { with a} total mass even larger than $10^{10}~M_\odot$.
 
Twenty-seven cases are { shown}, see Table \ref{tab1}. We consider only { SMBHBs that are individually observable} by PTA-SKA, generating a timing residual above a given threshold ($50$, $5$, $1$ ns), and merging in less than $10$ years. For these sources, we provide the total number of observations in $10$ years in { $n_{10,10}$} units (we consider $f_{\mathit p} = 8\cdot 10^{-7}$ Hz  only). The resolvable SMBHBs are mostly situated at a { red shift} $0.2 < z < 1.5 $.

The paucity of the rates in Table \ref{tab1} is far from promising. Further, there is a spread of $7$ orders of magnitude between the most pessimistic model combined with a $50$ ns timing precision and the   
most optimistic model combined with a $1$ ns timing precision. 
{ For a $10$-year survey of SMBHBs merging within} $10$ years, the Cc scenario predicts $4.7 \cdot 10^{-2}$ observations at $1$ ns, { whereas the Aa scenario predicts} $3.5 \cdot 10^{-9}$ at $50$ ns. 
We emphasise that the values in Table \ref{tab1} include those SMBHBs which { would not} enter any SLI bandwidth, as they are too massive. 

But a different reading of the results is possible, { as outlined in} the abstract. 
Indeed, these numbers may be interpreted as {\it ideal} upper limits for sequential detection, when { assuming} i) an SLI detector with a low cut-off coincident to the high cut-off frequency of PTA-SKA, that is $f_{\mathit s} = f_{\mathit p}$; ii) a noiseless interferometer. { Under} these two conditions, no source would remain undetected by SLI; finally iii) PTA (or SKA) and SLI operating simultaneously for the same survey duration.  

Further, the rates in Table \ref{tab1} are approximately proportional to the time to { merger}  of the SMBHB, and { are obviously proportional} to the duration of the  survey, { so} we may { switch} to { $N_{y^{-2}}$}  units. Thus, when referring to the conditions i-iii) above, the probability of sequential detection at $1$ ns will never be better than $4.7 ~\cdot10^{-4}$ y$^{-2}$, that is per year to { merger}  and per year of survey (Cc), or $3.3 ~\cdot10^{-6}$ y$^{-2}$ (Aa), since a factor { of} $100$ stands between the two units.   

A third presentation of the results { may be proposed}, when searching { for} the ideally { minimum} number of years necessary { for} a single sequential detection. The number of years will be given by { $\sqrt{1/N_{y^{-2}}}$}. At $1$ ns, for the Cc scenario we get $46$ years, { versus $550$ years for Aa}.  

This is an extremely { optimistic, {\it i.e.} fairly unrealistic,} scenario for sequential detection, as it { implies hypotheses i) and ii)} above, iii) pulsar timing and SLI surveys { lasting} $46$ or even $550$ years, and iv) { a} precision of $1$ ns (conversely, the number of years { may be less}, when rates include { inspirals with} masses lower than $1 \cdot 10^8 M_\odot$; provided PTA-SKA is sensitive to lighter masses, the latter have { a} longer inspiral time). A space mission lasting for decades or centuries is out of the question, but such a constraint { does not} have not to be fulfilled literally. Indeed, if the mass of the SMBHB { can be evidenced} from pulsar timing, it is possible to predict when the SMBHB would be visible by a SLI. It would { then be} the task of the space agencies to launch a mission in time for this rendez-vous, { a few months or a few centuries} after observation by { radio astronomy}.       

It is worth { pointing out} that the spread among different model predictions { decreases} as timing precision improves. From  Table \ref{tab1}, we conclude that only the precision of $1$ ns timing residuals deserves to be retained { for further} analysis, { as the other precisions are} coupled to even more negligible chances. 

A caveat is that the population of the merging SMBHBs is 
constructed out of the Millennium Run. Thus, the
total coalescence rate is fixed to that predicted by the 
Millennium. The actual coalescence rate in the universe
is poorly constrained, and it is mostly determined in the
high mass, low { red shift} range (relevant to this study), by
counting galaxy pairs (Patton et al. 2002; Lin et al. 2004,2008,2010; Bell et al. 2006; Xu et al. 2012). { The rates predicted by
pair} counting are consistent with those predicted by
semi-analytic merger trees (Volonteri et al. 2003), 
and { by the} Millennium Run within a factor of a few (Sesana et al. 2008,2009). 
The numbers are, therefore, uncertain
by a factor of a few due to { different estimates} in the merger rate. 

Further, the true SMBHB population might not be perfectly described by current models, or { might} come from a completely unexplored physical mechanism (Volonteri 2011).
\vskip 10pt
\subsection{Optimal { sequential} detection rate}

We now wish to compute how the values in Table \ref{tab1} are affected by introducing an SLI low cut-off frequency $f_{\mathit s}$. The answer depends heavily  upon the ring-down frequency of the merging system, equation (\ref{eche}). 
Here, we consider three different remnant spins ($ a = 0,~0.9,~0.99$)
and two different low cut-off frequencies for SLI 
($f_{\mathit s} = 1$ or $3 \cdot 10^{-5}$ Hz). We infer the number of possible sequential detections, assuming $1$ ns residual precision. 
We obtain the values { reported} in Tables \ref{tab2},\ref{tab3}. They are meant as {\it optimal} sequential detection rates; {   the noise characteristics of a given SLI have been ignored}. Further, for each column it is assumed that {\it all} remnants acquire the same final spin, {\it i.e.}, the coalescences produce remnants solely of spin $a=0$, or $a=0.9$, 
or $a=0.99$. Larger rates are associated with higher values of spin. 

\begin{table}
\centering{
\caption{Optimal sequential detection rates}
\label{tab2}
\begin{tabular}{@{}lccc}
\hline
Model & All spins & All spins & All spins \\ 
 & at $a = 0$ & at $ a = 0.9$ & at $a = 0.99$ \\
\hline
Aa & $1.8 ~\cdot~10^{-4}$ & $2.9 ~\cdot~10^{-4}$ & $3.1 ~\cdot~10^{-4}$ \\  
Ab & $3.2 ~\cdot~10^{-4}$ & $8.0 ~\cdot~10^{-4}$ & $9.7 ~\cdot~10^{-4}$ \\
Ac & $8.7 ~\cdot~10^{-4}$ & $2.5 ~\cdot~10^{-3}$ & $3.1 ~\cdot~10^{-3}$ \\
Ba & $5.33~\cdot10^{-4}$ & $1.18~\cdot10^{-3}$   & $1.4 ~\cdot~10^{-3}$ \\
Bb & $8.56~\cdot10^{-4}$ & $2.71~\cdot10^{-3}$   & $3.56~\cdot10^{-3}$ \\
Bc & $2.42~\cdot10^{-3}$ & $9.11~\cdot10^{-3}$   & $1.25~\cdot10^{-2}$ \\
Ca & $9.65~\cdot10^{-4}$ & $3.15~\cdot10^{-3}$   & $4.11~\cdot10^{-3}$ \\
Cb & $1.12~\cdot10^{-3}$ & $4.48~\cdot10^{-3}$   & $6.64~\cdot10^{-3}$ \\
Cc & $3.48~\cdot10^{-3}$ & $1.59~\cdot10^{-2}$   & $2.4~~\cdot10^{-2}$ \\
\hline
\end{tabular}}
\vskip5pt
\parbox[s]{8.7cm}
{\footnotesize NOTE.-- For the nine astrophysical scenarios, optimal (noiseless SLI) sequential detection rate{, { $n_{10,10}$} units,} or the total number of events in { $10$} years, for SMBHBs i) observed by PTA-SKA at $1$ ns level, ii) entering the SLI bandwidth, the $f_{\mathit s}$ cut-off frequency being at $10^{-5}$ Hz, iii) merging in less than $10$ years, iv) for a SLI survey duration of $10$ years. 
The SMBHBs are mostly situated at a { red shift} $0.2 < z < 1.5 $. All { the} remnants have identical spins (either $a = 0$, or $0.9$, or $0.99$).
}

\end{table}

\begin{table}
\centering{
\caption{Optimal sequential detection rates}
\label{tab3}
\begin{tabular}{@{}lccc}
\hline
Model & All spins & All spins & All spins \\ 
 & at $a = 0$ & at $ a = 0.9$ & at $a = 0.99$ \\
\hline
Aa & $3.8 ~\cdot10^{-6}$ & $7.1 ~\cdot10^{-5}$ & $1.1 ~\cdot10^{-4}$ \\  
Ab & $2.3 ~\cdot10^{-6}$ & $9.4 ~\cdot10^{-5}$ & $1.7 ~\cdot10^{-4}$ \\
Ac & $6.1 ~\cdot10^{-6}$ & $2.3 ~\cdot10^{-4}$ & $4.5 ~\cdot10^{-4}$ \\
Ba & $2.8 ~\cdot10^{-6}$ & $8.5 ~\cdot10^{-5}$ & $1.5 ~\cdot10^{-4}$ \\
Bb & $2.5 ~\cdot10^{-6}$ & $1.6 ~\cdot10^{-4}$ & $2.0 ~\cdot10^{-4}$ \\
Bc & $5.6 ~\cdot10^{-6}$ & $2.7 ~\cdot10^{-4}$ & $5.5 ~\cdot10^{-4}$ \\
Ca & $3.4 ~\cdot10^{-6}$ & $1.2 ~\cdot10^{-4}$ & $2.3 ~\cdot10^{-4}$ \\
Cb & $2.5 ~\cdot10^{-6}$ & $1.6 ~\cdot10^{-4}$ & $2.4 ~\cdot10^{-4}$ \\
Cc & $6.9 ~\cdot10^{-6}$ & $3.4 ~\cdot10^{-4}$ & $7.1 ~\cdot10^{-4}$ \\
\hline
\end{tabular}}
\vskip5pt
\parbox[s]{8.7cm}
{\footnotesize NOTE.-- Same caption as for Table \ref{tab2}, except for $f_{\mathit s} = 3 \cdot 10^{-5}$ Hz.}
\end{table}

The optimal { sequential} detection rates { of Tables \ref{tab2},\ref{tab3}} are, obviously, lower than the rates { of Table \ref{tab1}}, because most SMBHs have { a } ring-down 
frequency below the $f_{\mathit s}$ frequency cut-off. Numbers, in { $n_{10,10}$} units, are in the
range { $10^{-4}$ - $10^{-2}$}, if an { SLI} 
low frequency cut-off { at $10~\mu$Hz }is assumed. 

{ The chances of detection critically drop by} about two orders of magnitude { when the cut-off is shifted even slightly} to higher { frequencies}, from $10$ to $30$ $\mu$Hz. { Optimal sequential detection rates { are in the range $10^{-6}$ - $10^{-4}$} 
for a { SLI cut-off frequency at} $3\cdot 10^{-5}$ Hz. Further, we have tested that the rates are exactly zero in our models for a cut-off
{ at} $10^{-4}$ Hz. The spread in the rate values due to different spins lowers with a decreasing cut-off frequency.  

Again, the values may be read differently in { $N_{y^{-2}}$}  units, obtained by dividing { the values in Tables \ref{tab2} and  \ref{tab3} by a factor of $100$}. Finally, a single detection occurs every $79$ years for the Cc scenario, { every $587$ years} for the Aa { one}, both for a cut-off at $10$ $\mu$Hz and spin of $0.9$, while 
{ for a cut-off at $30$ $\mu$Hz and a spin of $0.9$, $542$ years are required for the Cc scenario, and $1187$ years for the Aa one}.
The preceding values { no longer imply} condition i) { in the previous section}, but the other conditions hold, { {\it i.e.}} ii) noiseless SLI, iii) pulsar timing and SLI surveys { lasting} $79$ or even $1187$ years, or alternatively { the possibility of launching an SLI mission at any time within a given period}, iv) $1$ ns timing precision, and v) all remnants possessing a spin of $0.9$.       

It is legitimate to ask what { the rates would be} for scenarios displaying multiple spins. Lousto et al. (2010a,b,c) show an analytic distribution for dry mergers following the Kumaraswamy (1980) functional form, peaked at $ a= 0.75$. For wet mergers, Dotti et al. (2010) provide a statistical distribution, peaked at $a=0.89$. { The distribution of cases between dry and wet mergers is not known}.


\subsection{Signal to noise-ratio (SNR) for NGO}}

We now turn to a specific configuration, namely NGO, and attempt to make a more realistic estimate by considering the contribution of noise. Computations relative to other SLI projects will be carried out when such { projects will have been} planned by space agencies. 
 
For the computation of the SNR for NGO\footnote{{ The contribution of confusion} noise from white dwarf binaries is marginal.} (Jennrich et al. 2011), we stick to the mass range of $2\cdot 10^{8-9}~M_\odot$, the dimensionless spin Kerr parameter ($a = 0, 0.9, 0.99$), and introduce four distances ($z = 0.2, 0.3, 1, 1.5 $), see Table \ref{tab4}.     
Recently, progress has been made on the computation of SNRs from inspiral to ring-down 
(Santamaria et al. 2009,2010; Ajith et al. 2011). 
Thus, the fifth column in Table \ref{tab4} shows the SNR for phenomenological { wave forms}, computed with the {\sc
  PhenomC} model (Santamaria et al.  2010).
For comparison and for the sole inspiral phase, the fourth column displays the SNR, computed using a post-Newtonian approximation (Poisson \& Will 1995). 

As expected, the computation of SNR for NGO shows that part of the sources that have come all the way from the PTA-SKA  bandwidth to that of NGO are not necessarily detected by the interferometer, due to the presence of noise. The heaviest SMBHB remnants at $z=0.3$ 
{ are not} detected if they are associated to a low spin, and for any spin value for $z > 0.5$, see Table \ref{tab4}. NGO is optimised for { lower mass} SMBHBs { and} this explains the low values of SNR for { the} sources { considered herein}.\\

\begin{table}
\centering{
\caption{SNRs for a NGO type configuration}
\label{tab4}
\begin{tabular}{@{}ccccc}
\hline
$ z $ & $ M (M_\odot )$   &   $ a $ & SNR                   & SNR           \\
      &               &             & { PN}                    & { {\sc PhenomC}} \\
     &               &              & Inspiral { only}         & { up to} ring-down\\
\hline
0.2 & $2~\cdot10^{8}$ & $ 0    $ & $1.08~\cdot10^{2}$ & $ 1.48~\cdot10^{3} $ \\
    &            & $ 0.9  $ & $1.08~\cdot10^{2}$ & $ 3.41~\cdot10^{3} $ \\
    &            & $ 0.99 $ & $1.08~\cdot10^{2}$ & $ 4.06~\cdot10^{3} $ \\
\cline{2-5}
    & $2~\cdot10^{9}$ & $ 0    $ & $0          $ & $ 1.08~\cdot10^{1} $ \\
    &            & $ 0.9  $ & $0          $ & $ 1.84~\cdot10^{2} $ \\
    &            & $ 0.99 $ & $0          $ & $ 2.86~\cdot10^{2} $ \\
\hline
0.3 & $2~\cdot10^{8}$ & $ 0    $ & $6.17~\cdot10^{1}$ & $ 8.73~\cdot10^{2} $ \\
    &            & $ 0.9  $ & $6.17~\cdot10^{1}$ & $ 2.02~\cdot10^{3} $ \\
    &            & $ 0.99 $ & $6.17~\cdot10^{1}$ & $ 2.41~\cdot10^{3} $ \\
\cline{2-5}
    & $2~\cdot10^{9}$ & $ 0    $ & $0          $ & $ 0           $ \\
    &            & $ 0.9  $ & $0          $ & $ 5.78~\cdot10^{1} $ \\
    &            & $ 0.99 $ & $0          $ & $ 9.12~\cdot10^{1} $ \\
\hline
1   & $2~\cdot10^{8}$ & $ 0    $ & $5.62       $ & $ 1.43~\cdot10^{2} $ \\
    &            & $ 0.9  $ & $5.62       $ & $ 3.35~\cdot10^{2} $ \\
    &            & $ 0.99 $ & $5.62       $ & $ 4.01~\cdot10^{2} $ \\
\cline{2-5}
    & $2~\cdot10^{9}$ & $ 0    $ & $0          $ & $ 0           $ \\
    &            & $ 0.9  $ & $0          $ & $ 0           $ \\
    &            & $ 0.99 $ & $0          $ & $ 0           $ \\
\hline
1.5 & $2~\cdot10^{8}$ & $ 0    $ & $0          $ & $ 7.1~~\cdot10^{1} $ \\
    &            & $ 0.9  $ & $0          $ & $ 1.68~\cdot10^{2} $ \\
    &            & $ 0.99 $ & $0          $ & $ 2.01~\cdot10^{2} $ \\
\cline{2-5}
    & $2~\cdot10^{9}$ & $ 0    $ & $0          $ & $ 0           $ \\
    &            & $ 0.9  $ & $0          $ & $ 0           $ \\
    &            & $ 0.99 $ & $0          $ & $ 0           $ \\
\hline
\end{tabular}}
\vskip5pt
\parbox[s]{8.7cm}
{\footnotesize NOTE.-- SNRs (angle-averaged, single mode $l=m=2$) relative to NGO for sources  located at $z = 0.2, 0.3, 1, 1.5$, of mass $2\cdot 10^{8-9}~M_\odot$, and of dimensionless spin Kerr parameter $a = 0,~0.9,~0.99$. The fourth column provides the { PN-}SNR for the inspiral phase { (Poisson \& Will 1995)}, while the last { column} the phenomenological SNR up to the ring-down { ({\sc PhenomC} model, Santamaria et al.  2010), using the conventions by } Berti et al. (2005,2006).}
\end{table}

\subsection{Discussion on PTA-SKA and NGO sequential detection}

The lowest frequency requirement on NGO (ESA 2011) { is $10^{-4}$ Hz}. { However}, the goal\footnote{Quoting ESA (2011) ...``The crucial difference between the requirement and the goal lies in the testing and verification procedures: performances are fully tested and verified against the requirements, whereas goals are observed only in terms of
design and analysis, {\it i.e.} the mission design must allow for measurements over the wider frequency band. The
distinction between goals and requirements is made to prevent excessive efforts on testing and verification, in
particular at low frequencies."...} 
of NGO is to reach $3\cdot 10^{-5}$ Hz (ESA 2011); further Jennrich et al. (2011) show that the sensitivity curve stretches below this frequency, down to $10^{-5}$ Hz. This sensitivity curve has been used for computation of the SNR, see Table \ref{tab4}.

{ To estimate the rates}, we must { look for} a commensurate answer to the spread of several orders of magnitude, due to different astrophysical scenarios (SMBH-host relation, accretion process; for the remnant, fast or slow spin, spin distribution {\it vis--\`a--vis} dry and/or wet mergers), and to the paucity of the events. Further, other uncertainties play an { important} role. Thus, a coarse estimate on the rates is adequate.  The other major uncertainties are the following. 

In Sect. $3.2$, we { showed} that a difference of $20$ $\mu$Hz in the SLI low frequency cut-off determines a difference in the rates (both { $n_{10,10}$} and  { $N_{y^{-2}}$}  units) of two orders of magnitude. Moreover, we { noted} that for a cut-off at $10^{-4}$ Hz 
the chances of sequential detection are null for our simulations. Will NGO { perform} according to its requirements or its goals? { The chances of}  sequential detection may be { meagre} or simply { nonexistant}. 

The timeline scenario is uncertain at both ends. When { will} the ambitious $1$ ns precision be achieved?  Will NGO fly or will { there be} another SLI-type mission? { And when? And of what duration?} Let us { assume} that both conditions, {\it i.e.} the residual precision requirement and an SLI launch, will be met in the second half of the next decade. Then, $5$ years may well represent the { maximum} value for $t_{\mathit r}$ that we can observe. When the latter is coupled to, e.g.,{ a $3$-year} SLI survey, this implies that the rates in { $n_{10,10}$} units, must be divided approximatively by a factor $(10/5)\cdot(10/3)= 6.6$. 

Finally, the interferometer noise will { further reduce} the rates. Altogether, it seems judicious to apply between one and two orders of magnitude of reduction to the rates of Table \ref{tab2}, if NGO goals on sensitivity are met and a cutoff at $10^{-5}$ Hz is assumed.    

\section{Conclusions}

We have studied SMBHBs { with a mass range} of $2 \cdot 10^8 - 2 \cdot 10^9~M_\odot$. The masses are normalised to a $(1+z)$ factor, the { red shift} $z$ lying between $z=0.2$ and $z=1.5$), where the individually detectable sources by PTA-SKA are expected to lie. For { a high PTA-SKA frequency cut-off of} $f_{\mathit p}= 8 \cdot 10^{-7}$ Hz, SMBHBs may pass from the pulsar timing to laser interferometry band in a period ranging from two months to eight years ($f_{\mathit p}= 4 \cdot 10^{-7}$ Hz, from thirteen months to fifty years). Furthermore, the source signals may be strong enough to be received by both { radio astronomical} and laser interferometric means. 

{ Unfortunately,} the astrophysical estimates act as the show-stoppers to sequential detection. Even a noiseless, extremely low frequency laser space interferometer, coupled to an highly performing pulsar timing of $1$ ns, will not allow to go beyond $4.7 ~\cdot10^{-4}$ y$^{-2}$, or $3.3 ~\cdot10^{-6}$ y$^{-2}$ sequential detections per year to { merger}  and per year of survey, for { the} most optimistic and pessimistic astrophysical { models}, respectively. 
We { are dealing with a} handful of individually detectable SMBHBs, and therefore it is no surprise that only a very tiny part of those sources merge within a time limit. 
Other factors { reduce the chances even further}: higher values of the SLI low cut-off frequency, low precision timing residuals, slowly rotating remnants, and { obviously SLI} noise.  
We have also found that the spread between the different astrophysical predictions { decreases} as timing precision improves, and  the SLI low cut-off frequency decreases. 

Given the paucity of rates, major changes may come from radically new astrophysical models, or from the achievement of sub-nanosecond precision (predictions may then converge to a rate of $n \cdot 10^{-1} y^{-2}$, per year to { merger}  and per year of survey). { For the models, the recent estimates on merger rates and gravitational waves amplitude by McWilliams et al. (2012) are encouraging.} 

Complementarity between PTA-SKA and SLI should not be dismissed { out of hand}. Approaching the two bands would be helpful, and efforts to increment the availability of observation time may create favourable conditions to enlarge the PTA bandwidth.
SLI should { remain interesting} for low frequencies in light of the proposed  higher frequency interferometers BBO (Phinney et al. 2003) and DECIGO (Seto et al. 2001). An SLI configuration having a cut-off frequency at $10^{-6}$ Hz was proposed by Bender (2004).

Opportunities may also lie in inverse search. Ring-down signals observed by SLI may trigger specific searches by { PTA-SKA}, and hopefully allow to dig out the SMBHB from the background in the { PTA-SKA} band.  
{ Another} opportunity
may be provided by observation of PTA-SKA and the
consequent evaluation of the SMBHB parameters (mass,
spin), leading to a prediction when the merger would
occur.

{ In the absence in the immediate future of} a detector covering the gap { between PTA-SKA and SLI frequency bands}, these efforts { might be rewarded}.  Indeed, { the} detection of even a single event would be of paramount importance for the understanding of SMBHBs and of the astrophysical processes connected to their formation and evolution.   

\section*{Acknowledgments}

A. Sesana (Golm) has contributed to the determination of ideal and optimal rates, while E. Berti ({ Mississipi}) to the computation of SNR. 
Discussions with M. Pitkin (Glasgow), I. Cognard and K. Liu (Orl\'eans), { and exchanges with} P. Bender (Boulder), are acknowledged. G. Mamon (Paris) is thanked for his interest in  the manuscript. 
\vskip 30pt    
\section*{Appendix}
\section*{ Scenarios}

\begin{figure} 
\includegraphics[width=9cm]{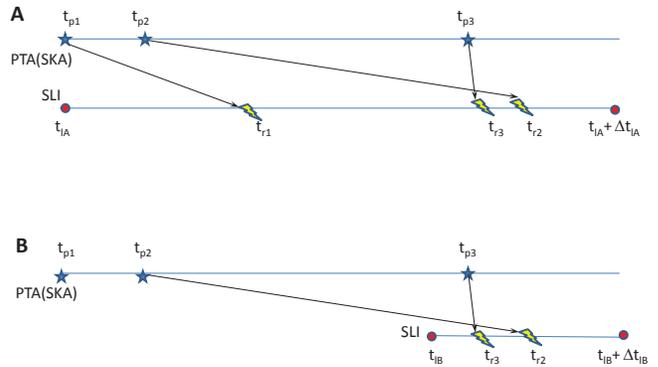}
\caption{Cases A and B for two different SLI lifetimes. The A case represents an SLI lifetime  of the same duration of the PTA-SKA survey, while for the B case, $\Delta t_{\mathit l}$ is shorter. 
{ The event at $t_{{\mathit p}1}$ is observed by PTA-SKA but not detected at $t_{{\mathit r}1}$ by SLI} in the B case. For the computation of rates, a reduction factor, given by $t_{{\mathit l}B} - t_{{\mathit l}A }$, is to be applied. } 
\label{fig3.eps}
\end{figure}

We define the following times:\\ 
- $t_{\mathit b,PTA-SKA}$ and $t_{\mathit e,PTA-SKA}$, the beginning and ending times of the PTA-SKA survey, respectively;\\ 
- $\Delta t_{\mathit PTA-SKA}$, the duration of the PTA-SKA survey;\\ 
- $t_0$, the time of the first observation by PTA-SKA;\\ 
- $t_{\mathit p}$, the exit time from the PTA-SKA band;\\ 
- $t_{\mathit r}$, the time when the SMBHB's remnant ring-down occurs (approximately the time of merging);\\ 
- $t_{\mathit s}$, the entry time in the SLI band;\\ 
- $t_{\mathit t} = t_{\mathit s} - t_{\mathit p}$, the transit time from { the} PTA-SKA to { the} SLI band; \\ 
- $t_{\mathit l}$, the launch date of the SLI mission;\\
- $\Delta t_{\mathit l}$, the SLI mission lifetime, that is the SLI duration survey, {\it e.g.}, $3$ years.

With the exception of {\it ideal} rates, we have assumed $t_{\mathit p}\neq t_{\mathit s}$, otherwise PTA-SKA and SLI would have contiguous bands. Further, $t_{\mathit r}\not\leq t_{\mathit s}$ for SMBHB entering the SLI band (otherwise $t_{\mathit s}$ is meaningless). Generally, inspiral of SMBHBs is such that   
$t_{\mathit p} \gg t_0 $. 
Sequential detection may occur if  
$t_{\mathit l} < t_{\mathit r} < t_{\mathit l} + \Delta t_{\mathit l}$, and for a given $t_{\mathit p}$, between two extremes: \\
i) $t_{\mathit r} \rightarrow t_{\mathit l+}$ for heavier SMBHBs;\\  
ii) $t_{\mathit r} \rightarrow (t_{\mathit l} + \Delta t_{\mathit l})_-$ for lighter SMBHB as they inspiral more slowly.

There is a large variety of operational schedules and {\it potential} events associated to PTA-SKA and the SLI mission(s), and to the launch date(s) of the latter. { We make no claim here to forecasting events, only to sizing} our analysis time wise. We attempt to reasonably group the scenarios into the following types: \\
I) $t_0$ and $t_{\mathit p}$ correspond to a PTA observation occurring in the second half of this decade; $t_{\mathit s}$ and $t_{\mathit r}$ to a SLI operating in the last quarter of { the} next decade; this leads to $t_{\mathit l}^{\mathit max}\sim 8-15$ y, and $(t_{\mathit l} + \Delta t_{\mathit l})^{\mathit max} \sim 11-18$ y.  \\   
II) $t_0$ and $t_{\mathit p}$ correspond to a PTA-SKA observation occurring in the first half of { the} next decade; $t_{\mathit s}$ and $t_{\mathit r}$ to a SLI operating in the last quarter of { the} next decade; this leads to $t_{\mathit l}^{\mathit max}\sim 3-10$ y, and $(t_{\mathit l} + \Delta t_{\mathit l})^{\mathit max} \sim 6-13$ y.  \\  
III) $t_0$ and $t_{\mathit p}$, $t_{\mathit s}$ and $t_{\mathit r}$ correspond to a PTA-SKA observation and to SLI operations both in the last quarter of { the} next decade; this leads to $t_{\mathit l}^{\mathit max}\sim 0-3$ y, and $(t_{\mathit l} + \Delta t_{\mathit l})^{\mathit max} \sim 3-6$ y. \\
IV) $t_0$ and $t_{\mathit p}$ correspond to a PTA-SKA observation occurring in the second half of { the} next decade; $t_{\mathit s}$ and $t_{\mathit r}$ to a {\it second generation} SLI operating in the 30s; this leads to $t_{\mathit l}^{\mathit max}\sim 0-15$ y, and $(t_{\mathit l} + \Delta t_{\mathit l})^{\mathit max} \sim 3-18$ y.  \\
V) Extremely large transit times may also be considered, either for space agencies maintaining in the future a sort of permanent presence  of SLI detectors, { as is} done nowadays in different bands of photon astronomy by successive launches. Alternatively, it may be {  conceivable to launch} an {\it ad hoc} SLI mission due to an alert provided by a PTA-SKA observation, { a} long time before. 

We conclude that the largest transit time $t_{\mathit t}^{\mathit max}$ for a SMBHB compatible { with the scenarios I-IV} is in the order of $20$ years, while an acceptable average is $10$ years. { This value is taken as the reference value} for our study. The shortest transit time $t_{\mathit t}^{\mathit min}$ depends upon the value of $f_{\mathit p}$, equation (\ref{transittime}), and it can be safely assumed as a couple of months.

{ Figure \ref{fig3.eps} shows how the SLI lifetime affects the rates of sequential detection. The rates in { $n_{10,10}$} and 
{ $N_{y^{-2}}$}  units imply 
a) $ t_{\mathit b,PTA-SKA} = t_{\mathit l}$ and b) $ t_{\mathit e,PTA-SKA} = t_{\mathit l} + \Delta t_{\mathit l}$. If the SLI lifetime is shorter than the { duration of the PTA-SKA} survey, the b) condition holds and the { $n_{10,10}$} rates must be multiplied by the factor 
$\Delta t_{\mathit PTA-SKA}\Delta t_{\mathit l}/100$, while the { $N_{y^{-2}}$}  rates by the factor $\Delta t_{\mathit PTA-SKA}\Delta t_{\mathit l}$.}


\section*{References}
{ 
 Ajith, P., et al. 2011, Phys. Rev. Lett., 206, 241101
}

{
 Amaro--Seoane, P., et al. 2012a, arXiv:1201.3621 [astro-ph.CO] 
}

{
 Amaro--Seoane, P., et al. 2012b, Class. Quantum Grav., 29, 124016 
}

 Antonucci, F., et al. 2012, Class. Quantum Grav., 29, 124014

  Bell, E.F., et al. 2006, Astrophys. J., 652, 270 

{ 
 Bender, P.L. 2004, Class. Quantum Grav., 21, 1203
}

  Berti, E., Buonanno, A. \& Will, C.M. 2005, Phys. Rev. D, 71, 084025 

  Berti, E., Cardoso, V. \& Will, C.M. 2006, Phys. Rev. D, 73, 064030

 Bertotti, B. 1984, in SPLAT: Space and Laser Applications and Technology, 25-30 March 1984 Les Diablerets (Noordwijk: ESA) ESA-SP-202, 147

 Blanchet, L., Spallicci, A. \& Whiting, B. 2011, Mass and motion in general relativity, Fundamental Theories of Physics, Vol. 162 (Berlin: Springer)

 Boyle, L. \& Pen, U.-L. 2010, arXiv:1010.4337 [astro-ph.HE]

 Burt, B.J., Lommen, A.N. \& Finn, L.S. 2011, Astrophys. J., 
730, 17

 Corbin, V. \& Cornish, N.J. 2010, arXiv:1008.1782 [astro-ph.HE] 

 Deng, X. \& Finn, L.S. 2011, Mon. Not. R. Astron. Soc., 414, 50

{
 Dotti, M., et al. 2010, Mon. Not. R. Astron. Soc., 402, 682
}

 Echeverria, F. 1989, Phys. Rev. D, 40, 3194


{
 ESA 2011, {\it NGO Revealing a hidden Universe:
opening a new chapter of discovery}, ESA/SRE(2011)19, December 2011
}

{
 Faller, J.E., et al. 1985, in Colloquium Kilometric Optical Arrays in Space, 23-25 October 1984 Carg\`ese (Noordwijk: ESA) ESA SP—226, 157

 Faller, J.E., et al. 1989, Adv. Sp. Res., 9, 107
} 

 Finn, L.S. \& Lommen, A.N. 2010, Astrophys. J., 
718, 1400

Forward,  R.L. \& Berman, D. 1967, Phys. Rev. Lett., 18, 1071

 Graham, A.W., Onken, C.A., Athanassoula, E. \& Combes, F. 2011, Mon. Not. R. Astron. Soc., 412, 2211   

{ 
 Grassi-Stini, A.M., Strini, G. \& Tagliaferri, G. 1979, Lett. N. Cim., 24, 212
}

 Gultekin, K., et al. 2009, Astrophys. J, 698, 198

 H\"aring, N. \& Rix, H.-W. 2004, Astrophys. J., 604, L89

 Hughes, S.A. 2002, Mon. Not. R. Astron. Soc., 331, 805

 Jenet, F.A., Lommen, A., Larson, S.L. \& Wen, L. 2004, Astrophys. J., 606, 799

 Jennrich, O., Petiteau A. \& Porter, E. 2011, Final configuration of the ELISA(NGO)
detector for science performance studies, v. 1.2,  
{\scriptsize https://lisa-light.aei.mpg.de/lisa-light/pub/DetectorConfigurations/FinalConfiguration/ELISA-NGO\_FinalConfig.pdf}
21 September 2011

{
 Kumaraswamy, P.  1980, J. Hydr., 46, 79
}

 Lauer, T.R., Tremaine, S., Richstone, D., \& Faber, S.M. 2007, Astrophys. J, 670, 249

Lee, K.J., et al. 2011, Mon. Not. R. Astron. Soc., 414, 3251 

 Lin, L., et al. 2004, Astrophys. J., 617, L9

 Lin, L., et al. 2008, Astrophys. J., 681, 232 

 Lin, L., et al. 2010, Astrophys. J., 718, 1158

{
 Liu, K., et al.  2011, Mon. Not. R. Astron. Soc., 417, 2916  
}

 Lommen, A.N. \& Backer, D.C. 2001, Astrophys. J., 562, 297 

{ 
 Lousto, C.O., Campanelli, M., Zochlower, Y. \& Nakano, H. 2010a, Class. Quantum Grav., 27, 114006

 Lousto, C.O., Nakano, H., Zochlower, Y. \& Campanelli, M. 2010b, Phys. Rev. D, 81, 084023; {\it ibidem} 2010c, Phys. Rev. D, 82, 129902(E) 
}

McWilliams, S.T., Ostriker, J.P. \& Pretorius, F., 2012 arXiv:1211.4590 

Mingarelli, C.M.F., Grover, K., Sidery, T., Smith, R.J.E. \& Vecchio, A. 2012, Phys. Rev. Lett., 109, 081104

 Patton, D.R., et al. 2002, Astrophys. J., 565, 208 

 Peters, P.C. 1964, Phys. Rev., 136, B1224

 Peters, P.C. \& Mathews J., 1963, Phys. Rev., 131, 435 

 Petiteau, A., Babak, S. \& Sesana, A. 2011, Astrophys. J., 732, 82 

{ 
Phinney, E.S., et al. 2003, {\it Beyond Einstein: from the Big Bang to black holes}, NASA Publication
NP-2002-10-510-GSFC
}

 Pierro, V. \& Pinto, I. 1996, N. Cim. B, 111, 631

 Pierro, V., Pinto, I.M., Spallicci, A.D., Laserra, E., \& Recano, F. 2001, Mon. Not. R. Astron. Soc., 325, 358

 Pitkin, M. 2012, Mon. Not. R. Astron. Soc., 425, 2688

 Pitkin, M., et al., 2008, J. Phys. Conf. Ser., 122, 012004

{ 
 Poisson, E.\& Will, C.M. 1995, Phys. Rev. D, 52, 848
}

 Rezzolla, L. 2009, Class. Quantum Grav., 26, 094023

{ 

 Santamar\'ia, L., Krishnan, B. \& Whelan, J.T. 2009, Class. Quantum Grav., 26, 114010

 Santamar\'ia L., et al. 2010, Phys. Rev. D, 82, 064016

}

 Sesana, A., Vecchio, A., \& Colacino, C. N. 2008, Mon. Not. R. Astr. Soc., 390, 192

Sesana, A., Vecchio, A. \& Volonteri, M. 2009, Mon. Not. R. Astron. Soc., 394, 2255

 Sesana, A. \& Vecchio, A. 2010, Class. Quantum Grav., 27, 084016

{
 Seto, N., Kawamura, S. \& Nakamura, T. 2001, Phys. Rev. Lett., 87, 221103
}

 Springel, V., et al. 2005, Nat., 635, 429

Tremaine, S., et al. 2002, Astrophys. J., 574, 740 

 Tundo, E., Bernardi, M., Hyde, J.B., Sheth, R.K. \& Pizzella, A. 2007, Astrophys. J.,
663, 57

 Volonteri, M. 2011, Journ\'ees LISA-France, Laboratoire APC AstroParticules et Cosmologie Paris, 9-10 May 2011,  {\scriptsize http://www.apc.univ-paris7.fr/LISA-France/2011-2012\_files/Volonteri.pdf}
 
 Volonteri, M., Haardt, F. \& Madau, P. 2003, Astrophys. J., 582, 559

 Xu, C.K., 2012, Astrophys. J., 747, 85

 Yardley, D.R.B., et al. 2010, Mon. Not. R. Astron. Soc., 407, 669


\end{document}